\documentclass[aps,prb,reprint,superscriptaddress,showkeys,notitlepage]{revtex4-2}
\usepackage{amssymb}
\usepackage{graphicx}
\usepackage[pdfusetitle,
 bookmarks=true,bookmarksnumbered=false,bookmarksopen=false,
 breaklinks=false,pdfborder={0 0 0},pdfborderstyle={},backref=false,colorlinks=true]
{hyperref}

\begin{document}

\title{Electron- and hole-doping on ScH$_{2}$ and YH$_{2}$: Effects on  
superconductivity without applied pressure}

\author{S. Villa-Cort{\'e}s}
\email{svilla@ifuap.buap.mx}

\author{O. De la Pe{\~n}a-Seaman}
\affiliation{Instituto de F{\'i}sica, Benem{\'e}rita Universidad Aut{\'o}noma de Puebla, Apartado Postal J-48, 72570, Puebla, Puebla, M{\'e}xico }

\date{\today}

\begin{abstract}
We present the evolution of the structural, electronic, and lattice dynamical properties, as well as the electron-phonon coupling and superconducting critical temperature ($T_c$) of ScH$_2$ and YH$_2$ metal hydrides solid solutions, as a function of the electron- and hole-doping content. The study was performed within the density functional perturbation theory, taking into account the effect of zero-point energy through the quasi-harmonic approximation, and the solid solutions Sc$_{1-x}M_{x}$H$_{2}$ ($M$=Ca,Ti) and Y$_{1-x}M_{x}$H$_{2}$ ($M$=Sr,Zr) were modeled by the virtual crystal approximation. We have found that, under hole-doping ($M$=Ca,Sr), the ScH$_2$ and YH$_2$ hydrides do not improve their electron-phonon coupling properties, sensed by $\lambda(x)$. Instead, by electron-doping ($M$=Ti,Zr), the systems reach a critical content $x \approx 0.5$ where the latent coupling is triggered, increasing $\lambda$ as high as $70\%$, in comparison with its $\lambda(x=0)$ value. Our results show that $T_c$ quickly decreases as a function of $x$ on the  hole-doping region, from $x=0.2$ to $x=0.9$, collapsing at the end. Alternatively, for electron-doping, $T_c$ first decreases steadily until $x=0.5$, reaching its minimum, but for $x > 0.5$ it increases rapidly, reaching its maximum value of the entire range at the  Sc$_{0.05}$Ti$_{0.95}$H$_2$ and  Y$_{0.2}$Zr$_{0.8}$H$_2$ solid solutions, demonstrating that electron-doping can improve the superconducting properties of
pristine metal hydrides, in the absence of applied pressure.
\end{abstract}


\maketitle

\section{Introduction}

In the last decade, several theoretical predictions have been made
on the crystal structure of stoichiometric and hydrogen-rich materials
at high pressures for which their electronic, dynamic and 
electron-phonon (el-ph) coupling
properties had been calculated\citep{cor6,Duan2019,PhysRevB.96.100502,BI2019}.
As a result from those predictions, several metal hydrides have been
proposed as conventional-superconductor candidates with a superconducting
critical temperature ($T_{c}$) near room 
temperature\citep{10.1038/srep06968,Liu6990,Wang6463}.
However, since 2015, the search for high-temperature superconductors has 
experienced a renewal, due to the discovery of 
phonon-mediated superconductivity on H$_{3}$S, with a $T_{c}$ of 203~K under 
pressures as high as 155 GPa\citep{203,eina}, as well as LaH$_{10}$ with 
$T_{c}$ in the range of 250-260 K\citep{La50,PhysRevLett.122.027001} at similar 
pressures, while, more recently, a $T_{c}$ of 262 K was measured in YH$_{9}$ at 182 
GPa\citep{PhysRevLett.126.117003}. 
Such breakthrough was achieved by a synergy between experiments and 
theoretical studies, which predicted the metallization and transition to the superconducting state of H$_{3}$S and LaH$_{10}$ at high 
pressure\citep{10.1038/srep06968,Liu6990,PhysRevLett.119.107001}, prior to the 
experimental verification.

The idea of hidrogen-rich materials metallized under pressure was first proposed by 
Ashcroft\citep{PhysRevLett.92.187002} and has become a hot topic for searching 
room-temperature superconductors. From the available experimental and 
theoretical data, it has been suggested that the propensity for 
superconductivity depends upon the species
used to build up the metal hydride, together with hydrogen. 
Some of the highest $T_c$ values are obtained from
hydrides constructed with elements that belong to the alkaline 
family as well as the first group of the transition metals (scandium group)\citep{BI2019}. Thus, several studies on the dynamical stability and superconducting state 
have been carried out in YH$_{n}$ and ScH$_{n}$ ($n\geqq3$) stoichiometric alloys. 

In particular, from calculations on YH$_{n}$, the $T_{c}$ was estimated around $305-326$~K at an applied pressure of 250 GPa for $n=10$ \citep{Liu6990,BI2019}; while for $n=6$ the $T_c$ values were on the range of $251-264$~K at 120 GPa\citep{BI2019}; and for $n=4$, lower values of $T_c$ ($84-95$~K) at 120 GPa were reported \cite{Yinwei}. However, was not until recently that measurements for a sibling system, YH$_{9}$, 
were realized, getting a $T_{c}$ of $262$~K  at 182 GPa\citep{PhysRevLett.126.117003,BI2019}. 
For the case of the ScH$_{n}$ family, high $T_{c}$ values in the range of $120-169$~K
were predicted for different members, like ScH$_{6}$, ScH$_{7}$, ScH$_{9}$, ScH$_{10}$, and ScH$_{12}$, all above of an applied pressure of 250~GPa\citep{Ye2018HighHO,BI2019}.

On both families, ScH$_{n}$ and YH$_{n}$, their members with lowest hydrogen content 
($n=3,2$) have not been so widely studied, mainly for their lack of compelling superconducting 
properties. For ScH$_{3}$ and YH$_{3}$, their highest $T_{c}$ have been estimated to be 20 and 40 K, respectively, at low pressure ($\approx$ 20 GPa)\citep{Kim2793,BI2019}.  For YH$_{2}$ there are not reports about its superconducting properties or $T_c$, so far. 
Regarding ScH$_{2}$, there are some theoretical studies that found that  $T_{c}$ rises under pressure, with a maximum value of 38 K at 30 GPa. 
Such behaviour was attributed to the hybridization between 
$1s$-states of H-atom and $d$-states of Sc-atom under pressure\citep{C6RA11862C,BI2019}.

Besides applied pressure, doping is another procedure to induce or
increase superconductivity by enhancing some properties like the electronic
density of states at the Fermi level ($N(0)$) or the electron-phonon
coupling. An example of such procedure is the one of Zhang \textit{et
al.}\citep{Zhang_2007}, where the substitution of Li by Be, Mg,
or Ca in LiH was studied. There, the dopant acts as a donor which delivers electrons
to the system, obtaining a $n$-doped material with a $T_{c}=$ 7.78
K for an electron content as high as 2.06, calculated at ambient pressure. 
More recently, Olea-Amezcua \textit{et al.}\citep{PhysRevB.99.214504}
shown the metallization of alkali-metal hydrides LiH, NaH, and KH
by doping with alkaline-earth metals Be, Mg, and Ca, respectively, and analyzed the 
superconducting properties as a function of concentration. The maximum 
estimated $T_{c}$ values were 2.1 K for Li$_{0.95}$Be$_{0.05}$H,
28 K for Na$_{0.8}$Mg$_{0.2}$H, and even 49 K for K$_{0.55}$Ca$_{0.45}$H, in absence of applied pressure. 
Such scheme to induce metallization and superconductivity on metal-hydrides 
could work as an alternative to the applied-pressure approach on the ScH$_{n}$ and YH$_{n}$.
So, in this paper we implement it  on the less-studied members of the family: ScH$_{2}$ and 
YH$_{2}$, and trace down the evolution of the structural, electronic and lattice dynamics  properties, as well as the el-ph coupling and $T_c$, as a function of concentration, inducing electrons ($n$-doped) and holes ($p$-doped) into the proposed systems.
Such approach is done by the construction of solid solutions with the metal atom of the hydride:
Sc$_{1-x}M_{x}$H$_{2}$ (M=Ca, Ti) and Y$_{1-x}M_{x}$H$_{2}$ ($M$=Sr, Zr)  within the Density
Functional Theory (DFT)\citep{PhysRev.140.A1133}, using the virtual crystal approximation 
(VCA)\citep{vca}, which has been successfully applied on the study of doped superconductors  \citep{PhysRevLett.88.127001,PhysRevLett.93.237002,PhysRevB.93.224513,PhysRevB.79.134523,PhysRevB.99.214504}. 

The paper is organized as follows. The computational details that support our method are presented in Section II. In Section III.A we present our results related to the structural properties; while in Section III.B the electronic structure analysis is shown. The lattice dynamics is discussed in Section III.C; and electron-phonon and superconducting properties, as well as $T_{c}$, are shown in Section III.D. Last, our conclusions are presented in Section IV.

\section{Computational details}

In order to study the superconducting state on the proposed systems, the electronic structure,
the phonon dispersion and the electron-phonon coupling properties
were obtained for the $Fm\bar{3}m$ crystal structure, without applied pressure. We calculated the ground state properties within the
framework of Density Functional Theory (DFT)\citep{PhysRev.140.A1133},
while the lattice dynamics and coupling properties were obtained within the Density
Functional Perturbation Theory (DFPT) \citep{0953-8984-21-39-395502,RevModPhys.73.515},
both implemented in the QUANTUM ESPRESSO suit code \citep{0953-8984-21-39-395502}. The calculations were performed with a 80~Ry cutoff for the
 plane-wave basis, and a $24\times24\times24$ $k$-point
mesh. The Perdew-Burke-Ernzerhof (PBE) functional \citep{PhysRevLett.77.3865}
was employed to take into account the exchange and correlation contributions.

Complete phonon spectra were accessed by a Fourier interpolation
of dynamical matrices calculated on a $8\times8\times8$ $q$-point
mesh. Corrections due to quantum fluctuations at zero temperature, zero-point energy 
(ZPE) effects,
are estimated through the quasi-harmonic approximation (QHA)\citep{10.2138/rmg.2010.71.3,PhysRevB.99.214504}
using the calculated phonon density of states (PHDOS). Within this
approximation, the phonon contribution to the ground-state energy
is taken into account and a new structural optimization of each concentration $x$ can be performed.
Thus, the electronic structure, lattice dynamics and electron-phonon
properties, calculated with these lattice parameters, include ZPE corrections.

To gain more insight, we also calculated the phonon linewidths of the
$\vec{q}\nu$ phonon mode $\gamma_{\vec{q}\nu}$ arising from the
electron-phonon interaction given by\citep{PhysRevB.6.2577,PhysRevB.9.4733}
\begin{equation}
\gamma_{\vec{q}\nu}=2\pi\omega_{\vec{q}\nu}\sum_{\vec{k}nm}\left|g_{\vec{k}+\vec{q},\vec{k}}^{\vec{q}\nu,nm}\right|^{2}\delta\left(\epsilon_{\vec{k}+\vec{q},m}-\epsilon_{F}\right)\delta\left(\epsilon_{\vec{k},n}-\epsilon_{F}\right),
\end{equation}
where $g_{\vec{k}+\vec{q},\vec{k}}^{\vec{q}\nu,nm}$ are the matrix
elements of the electron-phonon interaction 
(calculated over a denser $48\times48\times48$ $k$-point mesh), 
$\epsilon_{\vec{k}+\vec{q},m}$
and $\epsilon_{\vec{k},n}$ are one-electron band energies, with band
index $m,n$, and vectors $\vec{k}+\vec{q},~\vec{k}$,
respectively, while $\omega_{\vec{q}\nu}$ is the phonon frequency for
mode $\nu$ at wave-vector $\vec{q}$.

\begin{table*}
\caption{
\label{tab:table1} Calculated volume (in $a_B^3$, where $a_B$ denotes the Bohr radius)
and bulk modulus (GPa), for the static and ZPE schemes of the pristine metal hydrides and its respective percentage difference with respect to the experimental available data.
}

\begin{ruledtabular}
\begin{tabular}{ccccccc}
System & \multicolumn{3}{c}{Volume ($a_B^3$)} & \multicolumn{3}{c}{Bulk modulus (GPa)}\tabularnewline
\hline 
 & Static & ZPE & Other works & Static & ZPE & Other works\tabularnewline
\hline 
ScH$_{2}$ & 182.44 (0.98\%) & 190.44 (3.35\%) & %
\begin{tabular}{c}
184.25\footnote{Exp. Ref. \onlinecite{ScH2_exp}}, 183.67\footnotemark[2]\tabularnewline
179.77\footnotemark[3]\tabularnewline
\end{tabular} & 89.04 & 80.59 & %
\begin{tabular}{c}
91.73\footnote{Cal. Ref. \onlinecite{YANG2013119}}, 109.2\footnote{Cal. Ref. \onlinecite{Wolf_2000}}\tabularnewline
114\footnote{Exp. (300 K) Ref. \onlinecite{LITYAGINA199269}}, 98\footnotemark[4]\tabularnewline
\end{tabular}\tabularnewline
YH$_{2}$ & 238.21 (0.42\%) & 244.69 (3.15\%) & %
\begin{tabular}{c}
237.21\footnotemark[1], 238.45\footnotemark[2]\tabularnewline
238.38\footnote{Exp. 295 K Ref. \onlinecite{PhysRevB.45.10907}}, 237.65\footnote{Exp. 95 K Ref. \onlinecite{PhysRevB.45.10907}}\tabularnewline
\end{tabular} & 82.31 & 75.19 & 79.98\footnotemark[2], 101.1\footnotemark[3]\tabularnewline
\end{tabular}
\end{ruledtabular}

\end{table*}

The isotropic Eliashberg spectral function, $\alpha^{2}F\left(\omega\right)$,
is described as 
\begin{equation}
\alpha^{2}F\left(\omega\right)=\frac{1}{2\pi\hbar N\left(0\right)}\sum_{\vec{q}\nu}\delta\left(\omega-\omega_{\vec{q}\nu}\right)\frac{\gamma_{\vec{q}\nu}}{\omega_{\vec{q}\nu}},
\end{equation}
where $N\left(0\right)$ is the electronic density of states, per atom
and spin, at $\epsilon_{F}$. The average electron-phonon coupling constant $\lambda$, which quantifies the
coupling strength as well as the Allen-Dynes characteristic phonon frequency $\omega_{ln}$\citep{PhysRevB.12.905}, are related to the Eliashberg function as 
\begin{equation}
\lambda=2\int_{0}^{\infty}d\omega\frac{\alpha^{2}F\left(\omega\right)}{\omega}=\frac{1}{2\pi\hbar N\left(0\right)}\sum_{\vec{q}\nu}\frac{\gamma_{\vec{q}\nu}}{\omega_{\vec{q}\nu}^{2}},
\end{equation}
and 
\begin{equation}
\omega_{ln}=\exp\left\{ \frac{2}{\lambda}\int_{0}^{\infty}d\omega\frac{\alpha^{2}F\left(\omega\right)}{\omega}\ln\omega\right\}.
\end{equation}
Finally, the superconducting transition temperature $T_{c}$ was estimated
for each case by solving numerically the isotropic Migdal-Eliashberg gap equations\citep{Eliashberg,Bergmann1973,VILLACORTES2018371}, using the respective
$\alpha^{2}F\left(\omega\right)$ for each content $x$, and treating the Coulomb pseudopotential as an adjusted parameter.

\section{Results and discussion}

\subsection{Structural properties}

We performed structural optimizations of the cubic fluorite structure
 ($Fm\bar{3}m$ space group) with a primitive cell of three atoms (one
metal and two hydrogens) for the two alloy systems at different values
of metal content ($x$). Our volume results for the pristine 
metal dihydrides, ScH$_2$ and YH$_2$, under the ZPE and static schemes are in good agreement
with the experimental data\citep{PhysRevB.45.10907,YANG2013119},
as well as the bulk modulus results respect to other calculations on literature,
as we show in Table \ref{tab:table1}.

Regarding the solid solutions, for Sc$_{1-x}M_{x}$H$_{2}$ the equilibrium volume was 
determined for concentrations up to $x=0.95(0.9)$ of electron(hole) doping, while for 
Y$_{1-x}M_{x}$H$_{2}$, 
the range was for contents up to $x=0.8(0.9)$ of 
electron(hole) doping. 
Electron-doping thresholds were determined through dynamical instabilities, observed as 
imaginary frequencies in the phonon dispersion for larger content $x$. For hole-doping, although 
dynamical stability was found for the complete range ($0 \leq x \leq 1$), the system did not behave
as metallic for $x$ larger than the threshold.
It is important to mention that phonon instabilities in metal hydrides induced by alloying have been
observed before  \citep{doi:10.1063/1.4714549,PhysRevB.69.094205,PhysRevB.99.214504},
where such dynamical behavior have been linked to an increase of the heat of formation (i.e., the alloys become less stable).

\begin{figure}[b!]
\includegraphics[width=8.4cm]{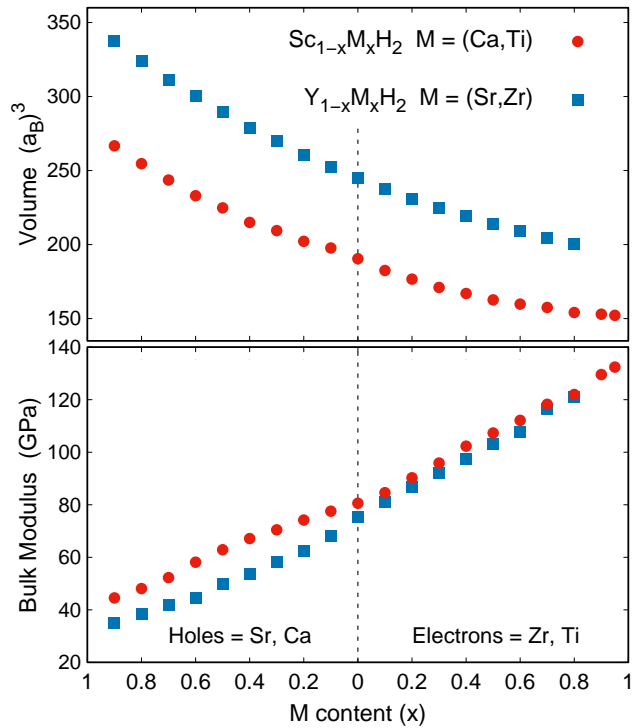}
\caption{\label{fig:bulk_vol}Volume and bulk modulus ($B_{0}$) for Sc$_{1-x}M_{x}$H$_{2}$
and Y$_{1-x}M_{x}$H$_{2}$ as a function of metal $M$ content ($x$).}
\end{figure}

In Fig. \ref{fig:bulk_vol}, we show the evolution of the volume and the bulk modulus ($B_{0}$)
as a function of metal $M$ content $x$. In both alloys, Sc$_{1-x}M_{x}$H$_{2}$
and Y$_{1-x}M_{x}$H$_{2}$,  increasing the electron content leads
to a monotonous reduction of the volume and an increase of $B_0$, while for hole doping such 
tendencies are opposite. This behavior indicates a strengthening of the
chemical bonding as the electron-content is increased, given by the increment
of Zr and Ti content, as well as a weakening of it as the hole-content grows as a
result of the increase of Sr and Ca content on the corresponding solid solutions.

\begin{figure}
\includegraphics[width=8.4cm]{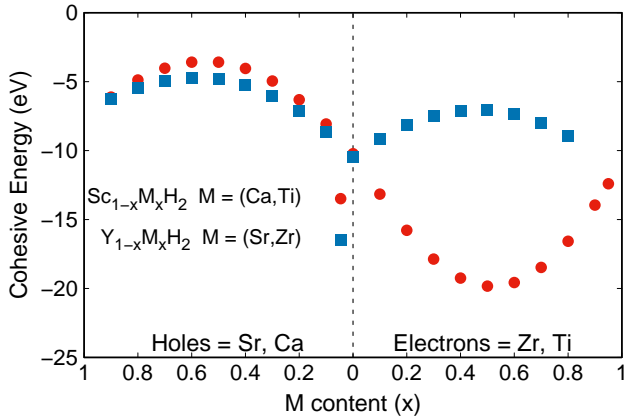}
\caption{\label{fig:cohe}Cohesive energy for Sc$_{1-x}M_{x}$H$_{2}$ and
Y$_{1-x}M_{x}$H$_{2}$ as a function of the metal $M$ content ($x$).}
\end{figure}
In Fig.\ref{fig:cohe} we show our calculated cohesive energy ($E_{coh}$) for the two systems
within their respective electron and hole range. This quantity
is used to characterize  stability of alloys and solid solutions, and is given by

\begin{equation}
E_{coh}=E_{alloy}^{tot}-(1-x)E_{N}^{a}-xE_{M}^{a}-2E_{H}^{a},
\end{equation}

where $E_{alloy}^{tot}$ is the total energy of the $N_{1-x}M_{x}$H$_{2}$
alloy at  content $x$, while $E_{N}^{a}$, $E_{M}^{a}$, and $E_{H}^{a}$
are the calculated total energies of the  isolated atoms 
$N$ = Y, Sc; $M$ = Sr, Zr, Ca, Ti; and hydrogen,  respectively. 
In general, the two solid solutions are in the stability range (negative $E_{coh}$), 
independent if it is hole- or electron-doping. In particular, the hole-doped systems are 
less stable than the pristine ones ($x=0$) (for $E_{coh}<0$, the larger the $E_{coh}$
absolute value, the more stable the system is). For the case of electron-doped systems,
although Y$_{1-x}$Zr$_{x}$H$_{2}$ follows the same observed tendency 
than the hole-doped systems, we found that Sc$_{1-x}$Ti$_{x}$H$_{2}$ is more stable 
than the pristine one, indicating the possibility to synthesize such solid solutions experimentally.

With the optimized lattice parameter for each system at different content for their electron and hole doping regions, we proceeded to calculate their electronic and lattice dynamical properties. Furthermore, we are  presenting  results obtained by  the  ZPE scheme. While the ZPE effects on the electronic properties are hardly visible, comparing with the static scheme, on the lattice dynamical ones the general effect is a noticeable softening (around $3.6\%$ at most). This tendency comes mainly from the  unit cell expansion  as the ZPE contribution to the energy is taken into account.

\subsection{Electronic properties}

In order to evaluate the effects of increasing the electron- and hole-content
on the electronic properties of the solid solutions, we 
analyze the evolution of the electronic band structure and the density of states
at the Fermi level, $N\left(0\right)$. 

\begin{figure}
\includegraphics[width=8.4cm]{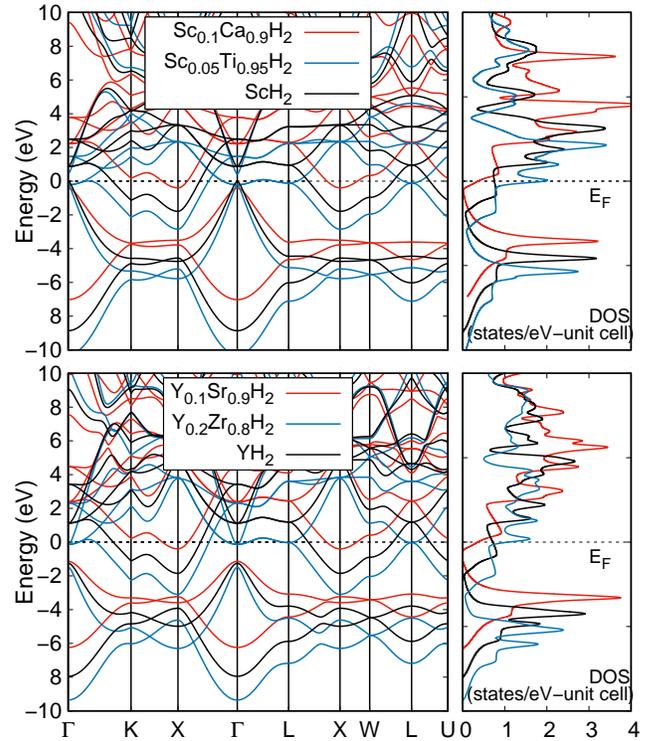}
\caption{\label{fig:bandas}Electronic band structure and density of states
(DOS), for Sc$_{1-x}M_{x}$H$_{2}$
and Y$_{1-x}M_{x}$H$_{2}$ at the pristine ($x=0$) and the threshold electron 
(Ti,Zr) and hole (Ca,Sr) doping levels.}
\end{figure}

In Fig. \ref{fig:bandas} we show the band structure for Sc$_{1-x}M_{x}$H$_{2}$ 
and Y$_{1-x}M_{x}$H$_{2}$ at the pristine and the threshold electron (Ti, Zr)
and hole (Ca, Sr) doping levels.
It can be seen that a twofold degenerated state, close to the Fermi
level, exists at the L-point in both pristine hydrides, which 
lies at 1.0~eV for ScH$_2$ and 1.2~eV 
for YH$_2$, giving place to a hole-like band at $E_F$.
As the electron-doping is increased, this band starts to fill up until a 
critical content, $x(\mbox{Ti}) \approx 0.75$ and $x(\mbox{Zr}) \approx 0.8$, where 
it is completely filled, and then changing its band character 
to electron-like for higher values of $x$. This indicates that an electronic topological transition (ETT) takes place, since the Fermi surface corresponding to the hole-like band disappears and a Fermi surface with electron character emerges. At $\Gamma$-point there is a threefold degenerated state above
the Fermi level at $0.9$~eV for ScH$_2$ and $1.1$~eV for YH$_2$, and as the electron-doping content increases on both systems, 
it shifts towards $E_F$. For Sc$_{1-x}$Ti$_{x}$H$_{2}$ the degeneracy of such state breaks at $x(\mbox{Ti})=0.5$, giving place to a new twofold state that continue 
to move towards $E_F$ as $x$ increases, crossing it at $x(\mbox{Ti})=0.7$, rising as an electron-like band. For Y$_{1-x}$Zr$_{x}$H$_{2}$, instead, the shift of the threefold state towards $E_F$ is monotonous,  crossing $E_F$ at $x(\mbox{Zr})=0.7$, creating an electron-like band. Such behavior, like the observed at L-point, also indicates and ETT at $\Gamma$-point.

\begin{figure}
\includegraphics[width=8.4cm]{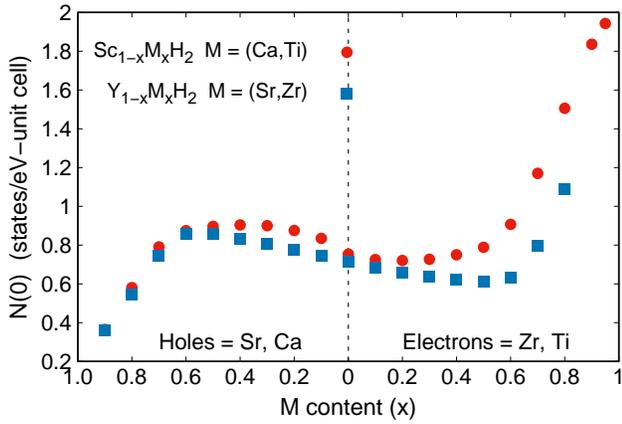}\caption{\label{fig:EFermi}
Evolution of the total density of states at the
Fermi level, $N\left(0\right)$, for Sc$_{1-x}M_{x}$H$_{2}$ and
Y$_{1-x}M_{x}$H$_{2}$ as a function of the $M$ content $x$.}
\end{figure}

Analyzing the evolution of the density of states at the Fermi level, 
$N\left(0\right)$, of both systems, Sc$_{1-x}M_{x}$H$_{2}$ and 
Y$_{1-x}M_{x}$H$_{2}$ (see Fig. \ref{fig:EFermi}), it can be observed that $N(0)$
presents little dispersion on the $M$-content range between the hole-doping of $x=0.6$ 
and electron-doping of $x=0.5$, for both hydrides. For higher $x$ on the hole-doping
regime, $N(0)$ reduces drastically, giving a maximum reduction of $\approx 60\%$ 
at the threshold content. At the electron doping regime, for $x \geq 0.5$, $N(0)$ shows an important  
increment of $\approx 60\%$ and $\approx 40\%$ for Sc- and Y-doped hydrides, 
respectively, indicating a steady improvement of the metallization with the increase  of electron doping. 

\subsection{Lattice dynamics} 

We now discuss the lattice dynamical properties as a function of doping
within the stability range of each solid-solution.

\begin{figure}[b!]
\includegraphics[width=8.4cm]{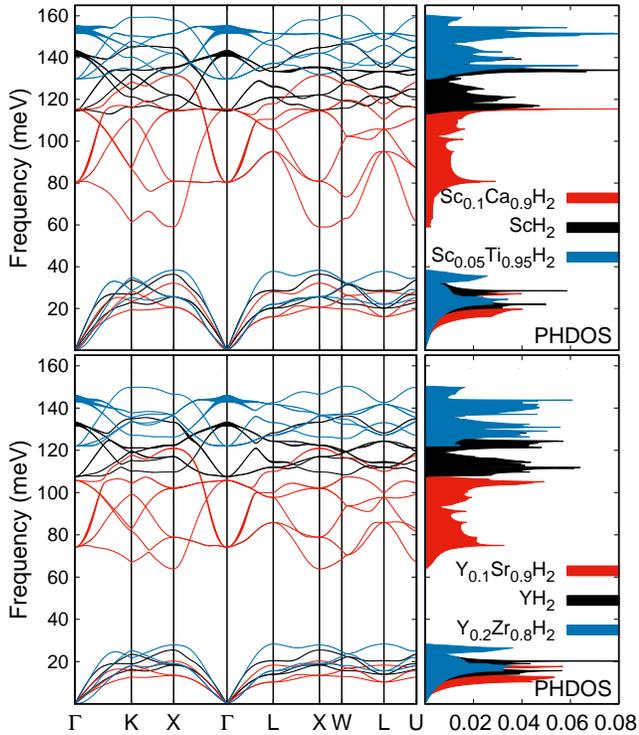}
\caption{\label{fig:phonons}  Phonon dispersion, 
linewidths (as vertical lines along the phonon branches) and PHDOS for 
Sc$_{1-x}M_{x}$H$_{2}$ and Y$_{1-x}M_{x}$H$_{2}$ at the pristine and the threshold electron 
(Ti,Zr) and hole (Ca,Sr) doping contents.}
\end{figure}

The phonon dispersion is presented on Fig. \ref{fig:phonons} 
including their respective phonon linewidth $\gamma_{\vec{q}\nu}$ and the phonon density
of states (PHDOS) for Sc$_{1-x}M_{x}$H$_{2}$ and Y$_{1-x}M_{x}$H$_{2}$, 
at the pristine $x=0$ and the threshold electron (Ti,Zr) and hole (Ca,Sr) doping contents. In general, for both systems,
the optical and acoustic branches soften as the hole-doping increases, while them are shifted to higher frequencies as the electron-doping rises. 
In particular for the optical branches, while them shift almost on a rigid way above the
frequencies of the pristine systems for the electron-doping cases, on the 
hole-doping solid-solutions such branches show, in addition to the softening, a renormalization mainly
localized along the K-X and W-L high-symmetry paths. Interestingly, the phonon linewidths $\gamma_{\vec{q}\nu}$ (vertical lines along 
the phonon branches) that are mainly localized around $\Gamma$ at the optical phonon
branches for $x=0$ increase its weight for the electron-doping regime, while 
it is reduced and depleted outside $\Gamma$ for the hole-doping, indicating a possible 
increment of electron-phonon coupling for the former, and a reduction for the later.

\subsection{Electron-phonon and superconducting properties}

\begin{figure}
\includegraphics[width=8.4cm]{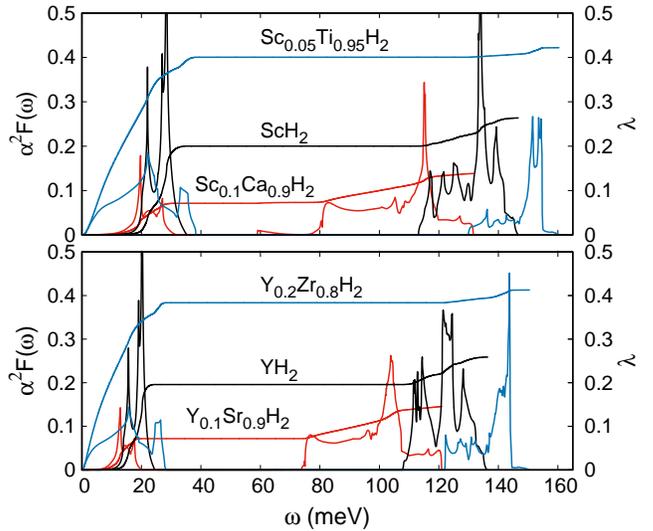}
\caption{\label{fig:A2f}Eliashberg function and the partial integrated electron-phonon coupling parameter $\lambda(\omega)$
for Sc$_{1-x}M_{x}$H$_{2}$ and Y$_{1-x}M_{x}$H$_{2}$ at
$x=0$ and at the threshold electron- and hole-doping content $x$ for each solid solution.}
\end{figure}

With the lattice dynamics information, the electron--phonon spectral functions 
$\alpha^{2}F\left(\omega\right)$ were calculated for the 
entire range of hole- and electron-doping stable regimes. As can be seen from $\alpha^{2}F\left(\omega\right)$ for the threshold 
electron- and hole-doping contents, as well as the pristine cases in Fig. \ref{fig:A2f},
as the electron(hole)-doping increases on both systems, 
the optical region of the Eliashberg function shifts to higher(lower) frequencies,
while the acoustical one gets wider(narrower). 

\begin{figure}
\includegraphics[width=8.4cm]{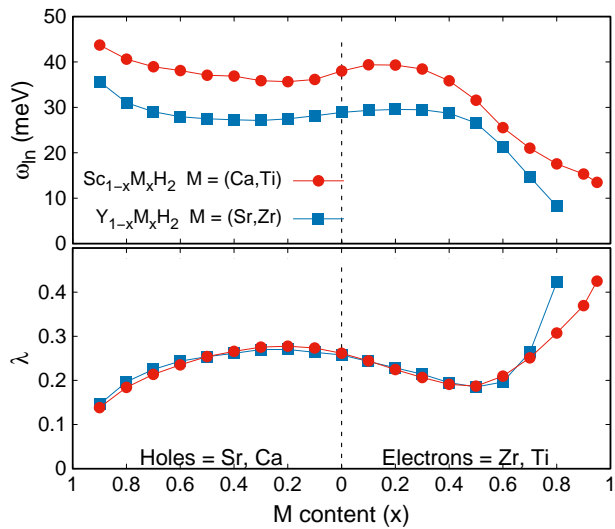}
\caption{\label{fig:lambda} The Allen-Dynes characteristic phonon frequency 
($\omega_{ln}$) and the electron--phonon coupling constant ($\lambda$) 
for Sc$_{1-x}M_{x}$H$_{2}$ and Y$_{1-x}M_{x}$H$_{2}$ as a function of the $M$ 
content ($x$).}
\end{figure}

As the Eliashberg spectral function determines the electron-phonon coupling parameter 
$\lambda$ (see Eq. 3), its evolution as a function of frequency, $\lambda(\omega)$, is 
shown in Fig.\ref{fig:A2f}. It can be observed that the main contribution to $\lambda$ 
comesfrom the the acoustic region, reaching almost its complete value if only such
frequency regime is taken into account. As the electron-doping content increases, the 
softening of the acoustic region boosts $\lambda$, leaving the optical 
one with a very minor effect. For the hole-doping case, the acoustic region loses weight, which
reduces $\lambda$, and despite of the softening of the optical region, this is not large
enough to increase its value.

In Fig. \ref{fig:lambda} the evolution of the average effective frequency 
$\omega_{\ln}(x)$ (see Eq. 4) 
and $\lambda(x)$ are presented for the studied solid solutions, as a function of 
$M$ content ($x$). For both solid solutions, $\omega_{\ln}(x)$ shows minor changes on the 
hole-doping regime. However, on the electron-doping region it starts to exhibit a 
decline at $x=0.4$, reaching the minimum value on the entire range (around 10~meV) at 
the threshold $x$ content on each case. $\lambda(x)$, 
for both systems, keeps fairly 
constant (around $\lambda(0)=0.25$) until $x=0.5$ on the hole-doping region, 
and then it starts to diminish as low as 0.14 at 
the limit value of hole-doping contents. For electron-doping, $\lambda(x)$ 
decreases slightly until 0.2 at $x=0.5$, and then it increases rapidly, reaching a 
maximum value of 0.42 at the threshold content on each solid solution. Thus, we have determined that, under hole-doping, the ScH$_2$ and YH$_2$ hydrides 
do not improve their electron-phonon coupling properties, showing a reduction of $44\%$ on $\lambda$. Instead,
by electron-doping, the systems reach a critical content $x \approx 0.5$ where the 
latent coupling is triggered, increasing $\lambda$ as high as $70\%$, in comparison 
with its value $\lambda(0)$ at the pristine systems.

\begin{figure}
\includegraphics[width=8.4cm]{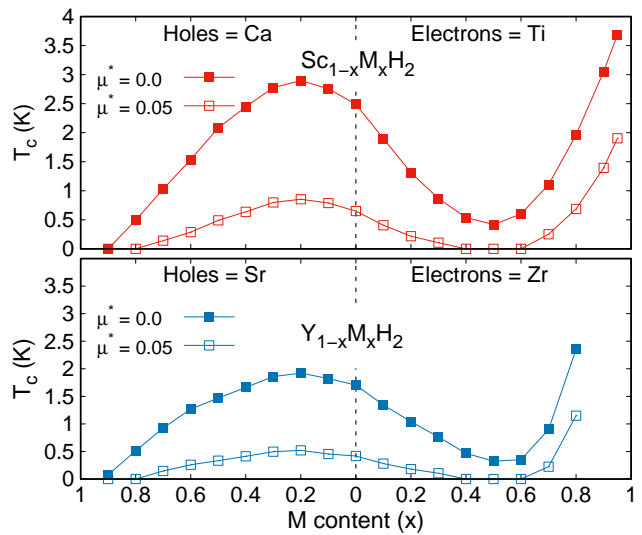}
\caption{\label{fig:TC}Calculated superconducting critical temperature $T_{c}$ as a function of  metal $M$ content ($x$) for 
Sc$_{1-x}M_{x}$H$_{2}$ and Y$_{1-x}M_{x}$H$_{2}$ at two different values of the Coulomb pseudopotential ($\mu^{*}=0,0.05$). 
} 
\end{figure}

Finally, the calculated electron-phonon coupling properties were used to obtain
estimates for the superconducting critical temperature $T_c$ as a function of 
content $x$ for both solid solutions. Numerically solving the isotropic
Migdal-Eliashberg gap equations, two different values of the Coulomb pseudopotential 
($\mu^{*}$) were employed: $\mu^{*}=0$, which provides an upper
limit for $T_{c}$, and $\mu^{*}=0.05$, which although is only a half of 
the typical value for many superconductors, it gives 
a more realistic estimate for $T_{c}$ and, at the same time, provides an idea 
of how strong $T_c$ can be affected by the variation of $\mu^{*}$. It can be seen from Fig. \ref{fig:TC}, for both solid solutions, that $T_{c}$ (calculated with $\mu^{*}=0$) shows a slightly increase at $x=0.2$ on the hole-doping region,  but then $T_{c}$ decreases quickly until $x=0.9$, where it collapses.
As a function of the electron-doping content, $T_{c}$ first
decreases steadily until $x=0.5$, reaching its minimum. For $x > 0.5$, $T_{c}$ increases rapidly, reaching a 
maximum $T_c$ value of $3.7$~K for Sc$_{0.05}$Ti$_{0.95}$H$_2$, and $2.4$~K for 
Y$_{0.2}$Zr$_{0.8}$H$_2$. Such behavior is basically the same when $T_c$ is calculated using $\mu^{*}=0.05$, 
however, now the maximum $T_c$ values are lower: $1.9$~K for the Sc-hydride solid 
solution, and $1.2$~K for the Y-hydride one. Despite the low $T_c$ values obtained, it was demonstrated that hole- and 
electron-doping could improve the superconducting properties of pristine metal hydrides,
in the absence of applied pressure. These results could help to design and implement 
novel schemes, additional to applied pressure, to increase the superconducting critical
temperature in other members of the ScH$_n$ and YH$_n$ hydrides.

\section{CONCLUSIONS}

We have performed a detailed analysis of the structural,
electronic, lattice dynamics, electron-phonon coupling, and superconducting properties of 
the metal-hydride solid-solutions 
Sc$_{1-x}M_{x}$H$_{2}$ ($M$=Ca,Ti) and Y$_{1-x}M_{x}$H$_{2}$ ($M$=Sr,Zr)  as a function of the electron- and hole-doping  content $x$. 
The evolution of the electronic band-structure, under electron-doping, indicates two 
different ETT that take place in both solid solutions: one 
at the L-point for $x(\mbox{Ti}) \approx 0.75$ and $x(\mbox{Zr}) \approx 0.8$; and
another at the $\Gamma$-point for $x \approx 0.7$ in both systems. 
While $N(0)$ is not improved in the whole hole-doping region, and even decreasing drastically for 
 $x \geq 0.6$, at the electron-doping 
regime for $x \geq 0.5$, however, $N(0)$ shows an important increment of approximately $60\%$ and $40\%$ for Sc- and Y-doped hydrides, respectively, 
indicating a steady improvement of the metallization. 
Simultaneously, the phonons soften by the hole-doping, whereas it harden 
for electron-doping. Interestingly, the linewidths
reduce for the former, and increase for the later, a behavior that indicates a suppression or
increment, respectively, 
of the electron-phonon coupling, which is confirmed by the evolution of $\lambda(x)$. 
In particular, under hole-doping, both hydrides do not improve $\lambda$, rather, its value drops approximately $44\%$ respect to its value at the pristine systems. Instead, by electron-doping, both solid-solutions reach a critical content $x \approx 0.5$ where the latent coupling is triggered, increasing 
$\lambda$ as high as $70\%$, in comparison with $\lambda(0)$. Then, due to all above, we found that for the hole-doping
region the superconducting critical temperature shows a slightly increase at $x=0.2$, but then it decreases quickly until $x=0.9$, where it collapses. For the electron-doping regime, $T_c$ reach a maximum value of $3.7(1.9)$~K for Sc$_{0.05}$Ti$_{0.95}$H$_2$, and 
$2.4(1.2)$~K for Y$_{0.2}$Zr$_{0.8}$H$_2$ taking $\mu^{*}=0.0(0.5)$. Then, our 
results shown that, in the absence of applied pressure, the hole- and electron-doping 
could improve the superconducting properties of pristine metal hydrides.

\begin{acknowledgments}
The authors thankfully acknowledge computer resources, technical advise,
and support provided by Laboratorio Nacional de Superc{\'o}mputo del Sureste
de M{\'e}xico (LNS), a member of the CONACyT National Laboratories. One
of the authors (S. Villa-Cort{\'e}s) also acknowledges the Consejo Nacional
de Ciencia y Tecnolog{\'i}a (CONACyT, M{\'e}xico) by the support under grant
769301. 
\end{acknowledgments}

\bibliographystyle{unsrt}
\bibliography{Art_MH2_rev4}

\end{document}